\documentclass[preprint,superscriptaddress,preprintnumbers,amsmath,amssymb]{revtex4-1}
\usepackage{graphicx}
\usepackage{amsmath}
\usepackage{gensymb}
\usepackage{mathtools}
\usepackage{comment}

\begin{document}
\title{Direct and parametric synchronization of a graphene self-oscillator}
\author{S. Houri} \email{s.houri@tudelft.nl}
\affiliation{Kavli Institute of Nanoscience, Delft University of Technology, Lorentzweg 1, 2628 CJ Delft, The Netherlands}
\author{S. J. Cartamil-Bueno}
\affiliation{Kavli Institute of Nanoscience, Delft University of Technology, Lorentzweg 1, 2628 CJ Delft, The Netherlands}
\author{M. Poot}
\affiliation{Kavli Institute of Nanoscience, Delft University of Technology, Lorentzweg 1, 2628 CJ Delft, The Netherlands}
\author{P. G. Steeneken}
\affiliation{Kavli Institute of Nanoscience, Delft University of Technology, Lorentzweg 1, 2628 CJ Delft, The Netherlands}
\author{H. S. J. van der Zant}
\affiliation{Kavli Institute of Nanoscience, Delft University of Technology, Lorentzweg 1, 2628 CJ Delft, The Netherlands}
\author{W. J. Venstra}
\affiliation{Quantified Air, Lorentzweg 1, 2628 CJ Delft, The Netherlands}

\date{\today} %Direct and Parametric entrainment of a graphene oscillator

\begin{abstract}
%This paper explores the synchronization dynamics in a single layer graphene nanomechanical oscillator at 3~K.%
We explore the dynamics of a graphene nanomechanical oscillator coupled to a reference oscillator. Circular graphene drums are forced into self-oscillation, at a frequency $\mathrm{f_{osc}}$, by means of photothermal feedback induced by illuminating the drum with a continuous-wave red laser beam. Synchronization to a reference signal, at a frequency $\mathrm{f_{sync}}$, is achieved by shining a power-modulated blue laser onto the structure. We investigate two regimes of synchronization as a function of both detuning and signal strength for direct ($\mathrm{f_{sync} \approx f_{osc}}$) and parametric locking ($\mathrm{f_{sync} \approx 2 f_{osc}}$). We detect a regime of phase resonance, where the phase of the oscillator behaves as an underdamped second-order system, with the natural frequency of the phase resonance showing a clear power-law dependence on the locking signal strength. The phase resonance is qualitatively reproduced using a forced van der Pol-Duffing-Mathieu equation.
\end{abstract}

\maketitle

\indent\indent\ Synchronization, also known as entrainment, is the phenomenon by which self-sustained oscillators mutually lock their frequencies and phase. %Synchronization could equally be achieved under the effect of an eternal reference signal when $\mathrm{f_{sync} \approx f_{osc}}$. %
Synchronization was first observed in a pair of coupled mechanical clocks by Huygens \cite{huygens1895letters,oliveira2015huygens}. Synchronized oscillators occur in a wide variety of engineered and biological systems such as injection-locked time keeping devices, the cardiac pacemaker cells and groups of fireflies  \cite{strogatz2003sync,matheny2014phase,bagheri2013photonic,pikovsky2003synchronization}. To study these phenomena experimentally, NanoElectroMechanical Systems (NEMS) have been proposed as representative model systems. Indeed, their strong nonlinearity, tunability, and convenient time scales make detailed experimental studies of synchronization possible, including the observation of features such as phase slipping, phase locking, phase inertia, and phase oscillation \cite{cross2004synchronization,shim2007synchronized,zhang2012synchronization,barois2014frequency,matheny2014phase}. Compared to top-down fabricated NEMS devices, graphene nanomechanical systems offer enhanced nonlinear response due to their extreme aspect ratio. This enables new experimental studies of parametric synchronization and phase-oscillation dynamics, which are the topic of this Letter.\\
\begin{comment}
power-dependence of the , and oscillate at high frequencies. These properties make it possible to access experimentally interesting dynamics give rise to an    thin are the ultima
In addition to direct synchronization, the oscillator can also lock to a reference signal with a frequency close to an integer multiple of the oscillator frequency. This case, where $\mathrm{f_{sync} \approx nf_{osc}}$, is known as parametric entrainment, and demonstrated in clamped-clamped beam nanomechanical systems and Josephson junction oscillators \cite{lin2014josephson}.\\

\end{comment}
\indent\indent In this work we demonstrate synchronization of a single-layer graphene (SLG) nanomechanical oscillator to an optical reference signal. Two cases are considered: synchronization to a reference frequency close to the oscillator frequency, and close to twice the oscillator frequency. We investigate the synchronization dynamics for both cases and demonstrate the presence of phase oscillations, and show that their frequencies exhibit a distinct power-law dependence on the strength of the reference oscillator. The phase oscillations, are explained using a van der Pol-Duffing-Mathieu equation, and are shown to occur when the nonlinear spring constant of the oscillator exceeds a threshold value.\\
\indent\indent\ The oscillator is fabricated by transferring a single layer of chemical vapor deposition (CVD) grown graphene onto an silicon substrate with circular cavities, which are etched into a 632~nm thick thermally grown silicon oxide layer. To reduce thermal drift, the graphene drum is placed in a cryogenic chamber with optical access (Montana Instruments), and cooled down to 3~K at a pressure of $\mathrm{<10^{-6}}$~mbar. Figure 1 shows the device and the setup. To induce self-oscillations, a red He-Ne laser ($\mathrm{\lambda}$ = 633~nm) is focused on the drum. The reflection from the silicon bottom of the cavity creates a partial standing wave which introduces a position-dependent thermally-induced mechanical tension in the structure \cite{barton2012photothermal}. The resulting photothermal force gradient, $\mathrm{\nabla F_{ph}}$, modifies the effective damping, given as $\mathrm{\Gamma_{eff} = \Gamma \left (1+\frac{\omega_0}{\Gamma} \frac{\omega_0 \tau}{1+\omega^2 \tau^2} \frac{\nabla F_{ph}}{\kappa} \right)}$ where, $\mathrm{\Gamma}$ ($\mathrm{\Gamma} = {\omega_0}/{Q}$) is the damping without feedback , $\mathrm{\omega_0}$ and $\mathrm{\kappa}$ are the natural frequency and spring stiffness of the graphene drum, and $\mathrm{\tau}$ is the thermal delay time \cite{metzger2004cavity,metzger2008optical}. The thickness of the oxide layer is chosen as to maximize $\mathrm{\nabla F_{ph}}$. As a result, the effective damping becomes negative at low laser power, and the drum enters a regime of self-oscillation.\\
\indent\indent The membrane's motion is detected using an interferometer as described in Refs. \cite{bunch2007electromechanical,castellanos2013single}. Briefly, a small portion of the incident red laser power is reflected off the graphene surface, and its interference with the light reflected from the silicon substrate underneath modulates the reflected intensity which is detected with a high-speed photodiode, as shown schematically in Fig.~1(a). The measurements are performed at an incident laser power of 10~mW. The motion of the graphene drum is recorded in the time-domain by sampling the photodiode output at 1~GS/s % during 1~ms ($\mathrm{\approx 15000}$ oscillations)%
 using an oscilloscope. % The time domain data of the oscillator and the reference signal are post-processed on a computer.%
 At the same time, an external reference signal, to which the graphene drum oscillator will be locked, is provided by a blue laser diode (2.5~mW, $\mathrm{\lambda}$ = 405~nm) whose intensity is electronically modulated. \\
\begin{comment}
\indent\indent The partial optical standing wave caused by the reflection of the He-Ne laser from the cavity bottom causes position-dependent heating of the graphene, which induces tension in the structure and results in a photothermal force gradient $\mathrm{\nabla F_{ph}}$. This modifies the effective damping which is now given as $\mathrm{\Gamma_{eff} = \Gamma \left (1+Q \frac{\omega_0 \tau}{1+\omega^2 \tau^2} \frac{\nabla F_{ph}}{\kappa} \right)}$, where $\mathrm{\Gamma}$ and $\mathrm{Q}$ are the mechanical damping and quality factor respectively, $\mathrm{\omega_0}$ and $\mathrm{\kappa}$ are the natural mode frequency and spring stiffness of the mechanical resonator, and $\mathrm{\tau}$ is the thermal delay time \cite{metzger2004cavity,metzger2008optical}. Depending on the wavelength and cavity depth, $\mathrm{\nabla F_{ph}}$ can be positive or negative, with the latter causing self-oscillation when the overall term inside the parenthesis is negative.\\
\end{comment}
\indent\indent\ Figure~1(b) shows the time-domain signals: the yellow trace indicates the free-running oscillator and the blue trace shows the output of the oscillator when the reference oscillator signal is applied. % with $\mathrm{f_{sync} = 15.19}$~MHz and $\mathrm{P_{d} = 0}$~dBm (corresponding to 1.5~mW modulation of the blue laser output).%
Figure~1(c) displays a zoom of the oscillations in more details. Figure~1(d) shows the power spectral densities (PSD) of the displacement signal, obtained by taking the FFT of the time traces. The spectral purity of the peak, given by its full-width at half-maximum (FWHM), is significantly better in the case the reference signal is applied (FWHM $\mathrm{< 1}$ kHz) compared to the case without the reference signal (FWHM $\mathrm{\approx 35}$ kHz). While this is an indication that the SLG drum motion is locked to the reference oscillator, the PSD does not provide information regarding the phase coherence.
\begin{figure}[bh]
	\graphicspath{{Figures/}}
	\includegraphics[width=85mm]{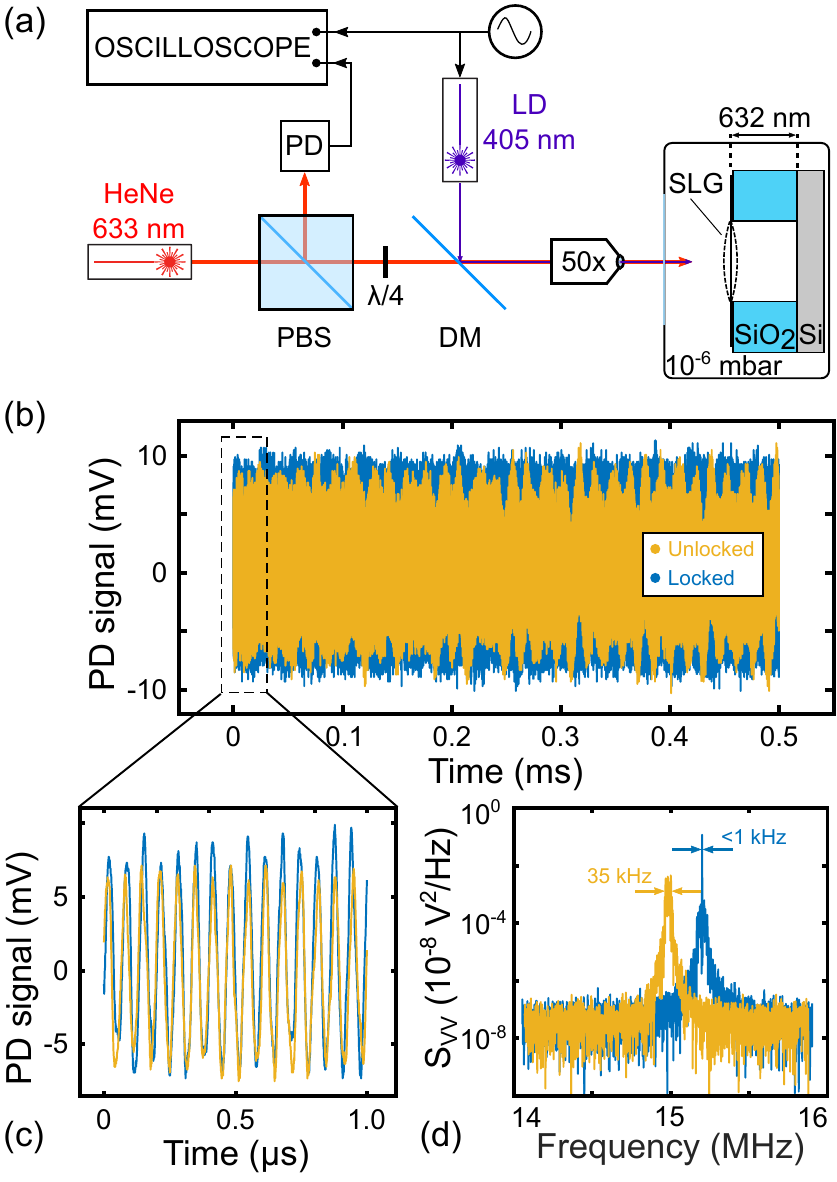}
	\caption{(a) Schematic representation of the measurement setup. A red He-Ne laser and a modulated blue laser are focused onto the drum via a window in the vacuum chamber of the cryostat at a temperature of 3 K. The displacement of the drum is detected using a photodiode (PD) and sampled with a digital oscilloscope. (b) A time-domain trace of the photodiode output for a free-running (yellow) and a synchronized (blue) oscillator. The frequency and power of the reference signal are $\mathrm{f_{sync} = 15.19}$~MHz and a modulation strength of $\mathrm{P_{d} = 1.5}$~mW respectively. (c) Zoom of the oscillation signal. (d) Power spectral density of the displacement and reference signals taken over a 1~ms time interval.}
\end{figure}
\indent\indent A more detailed picture of the oscillator phase is obtained by plotting the displacement signal on a slow (microseconds) and a fast (nanoseconds) time scale \cite{strogatz2003sync,barois2014frequency}. Figure~2(a) shows such a plot for the freely running oscillator, where the phase diffuses after a few hundred microseconds \cite{chen2013graphene}. In contrast, when the reference signal is applied (panel b) the phase is coherent during the measurement ($\mathrm{\sim}$1~ms). This demonstrates that the oscillator is synchronized to the reference signal. Interestingly, small phase fluctuations are noticeable on the slow time scale, which could indicate the presence of noise or higher order phase dynamics. These phase fluctuations become more apparent by plotting the in-phase component of the displacement versus its quadrature component with respect to the reference oscillator. The freely running oscillator Fig.~2(a) right panel shows a homogeneously distributed phase, while the locked oscillator phase (b, right panel) takes a fixed value. Note that a noise-free synchronized system would be represented by a single dot, significant fluctuations in both phase and amplitude are apparent in the synchronized graphene drum oscillator.\\
\indent\indent\ To explain the dynamics of the synchronized oscillator in the presence of noise, we describe the system using the Adler equation \cite{adler1946study,paciorek1965injection}:
\begin{equation}
	\mathrm{{\dot\phi} = -\frac{\mathrm{d}V(\phi)}{\mathrm{d}\phi} = - \Delta \omega + \gamma \sin (\frac{m}{n}\phi) + \xi(t)}.
	%\label{eqn:Eq2}
\end{equation}
\noindent
Here $\mathrm{V(\phi)}$ is a periodic potential, $\mathrm{\phi}$ is the phase difference between the graphene oscillator and the reference signal, $\mathrm{\gamma}$ is the amplitude of the reference signal, $\mathrm{\Delta \omega}$  is the detuning between the oscillator's natural frequency ($\mathrm{\omega_{osc} = 2\pi f_{0}}$) and the reference signal ($\mathrm{\omega_{sync} = 2\pi f_{sync}}$). $\mathrm{\xi(t)}$ is an additive stochastic term that represents the Brownian force noise. Synchronization occurs if $\mathrm{\Delta \omega = n \omega_{sync} - m \omega_{osc}}$, where m and n are integers. In the above experiment $\mathrm{m = n = 1}$, which results in direct synchronization. In the following section we also consider the case where $\mathrm{m = 2n = 2}$, which results in a higher order (parametric) synchronization \cite{adler1946study,pikovsky2003synchronization}.\\
\begin{figure}[h]
	\graphicspath{{Figures/}}
	\includegraphics[width=85mm]{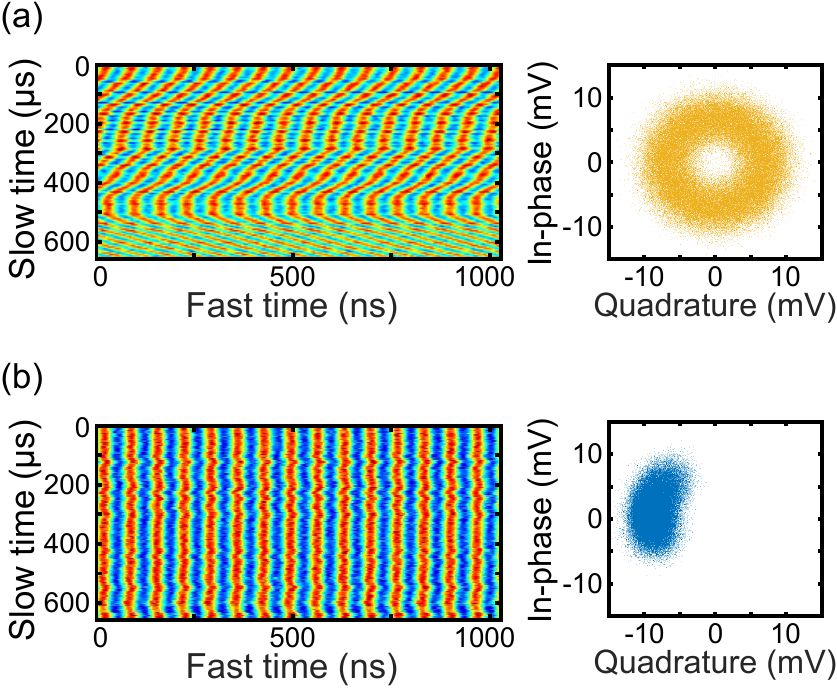}
	\caption{Raster plot (left panels) of the PD voltage of the free running (a) and locked (b) oscillators. The right panels show the corresponding I-Q plots.}
\end{figure}
\begin{figure*}[ht]
	\graphicspath{{Figures/}}
	\includegraphics[width=170mm]{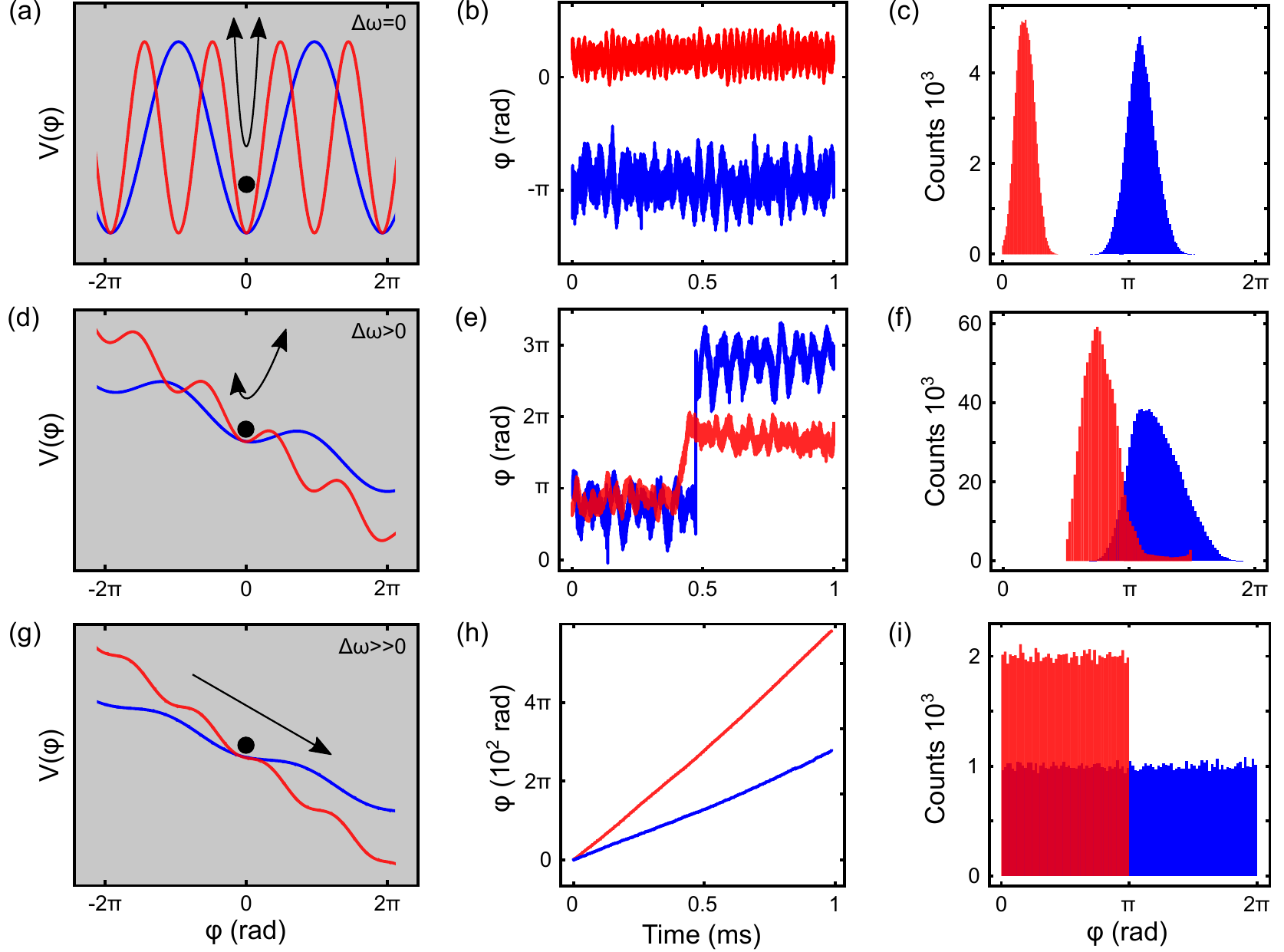}
	\caption{Washboard potential representation of entrainment, grey panels, for $\mathrm{\Delta \omega = 0}$ (a), $\mathrm{\Delta \omega \approx 0}$ (d), and $\mathrm{\Delta \omega >0}$ (g). shown for the direct (blue) and parametric cases (red). (b)  phase of the locked oscillator and the corresponding histogram (c). (e) noise-induced phase slips in a synchronized oscillator, and (f) the corresponding skewed-Gaussian distribution. Free-running phase of a unlocked oscillator (h) and the corresponding histogram showing a uniformly distributed phase (i).}
\end{figure*}
\begin{comment}

\indent\indent The above stochastic differential equation corresponds to a particle trapped in a potential well $\mathrm{V(\phi)}$, the potential can be obtained by integrating the deterministic terms in Eq.~(1) \cite{pikovsky2003synchronization} as a function of $\mathrm{\phi}$ to give: \\

\begin{equation}
\mathrm{V(\phi)} = - \Delta \omega + \gamma \sin (m\phi) + \xi(t),
%\label{eqn:Eq2}
\end{equation}

\end{comment}
\indent\indent Figure~3(a) shows the potential $\mathrm{V(\phi)}$ which has a period of $\mathrm{\frac{m}{n}2\pi}$. The blue curve represents the case where $\mathrm{f_{sync} = f_{osc}}$, while the red curve represents the parametric case with $\mathrm{f_{sync} = 2f_{osc}}$. The phase of the oscillator is trapped in the potential minimum and fluctuates under the effect of noise. Figure~3(b) shows the experimentally obtained phase difference, as calculated by taking the Hilbert transform of the measured time trace for %The phase difference $\mathrm{\phi}$ is obtained from the experimental data by taking the digital Hilbert transform. can be seen from experimental data in Fig.~3(b) for%
 $\mathrm{\Delta \omega =0}$. Here the direct forcing frequency $\mathrm{f_{d} = 15.19}$~MHz and the power $\mathrm{P_{d} = 1.1}$~mW, while for the parametric case the forcing frequency $\mathrm{f_{p} = 30.3}$~MHz and power $\mathrm{P_{p} = 1.5}$~mW. A slight detuning, $\mathrm{\Delta \omega\neq0}$, breaks the symmetry and causes the washboard potential to become tilted as shown schematically in Fig.~3(d) for $\mathrm{f_{d} = 15.01}$~MHz, $\mathrm{P_{d} = 0.75}$~mW, and $\mathrm{f_{p} = 30.265}$~MHz, $\mathrm{P_{p} = 0.35}$~mW. As the asymmetry created by tilting the potential reduces the barrier height, the system is now more prone to noise-induced phase slips where the phase undergoes a jump to the adjacent local minimum as the experimental data show in Fig.~3(e). Note that the direct forcing shows phase slips of $\mathrm{2\pi}$ whereas the parametric forcing shows phase slips of $\mathrm{\pi}$ as expected by theory. The asymmetry of the potential well is clearly reflected in the phase histograms. While a symmetric potential shows a Gaussian distribution Fig.~3(c), a tilted potential results in a skewed-Gaussian distribution Fig.~3(f). If the detuning is increased further, the tilt increases and the potential no longer represents a local minima, as shown in Fig.~3(g) for $\mathrm{f_{d} = 14.78}$~MHz, $\mathrm{P_{d} = 0.35}$~mW, and $\mathrm{f_{p} = 30.01}$~MHz, $\mathrm{P_{p} = 0.35}$~mW. The synchronization is lost, and the oscillator phase is free-running with respect to the reference signal, as shown in Fig.~3(h). In this case, the phase histogram is uniformly distributed over the  $\mathrm{2\pi}$ and $\mathrm{\pi}$ range, Fig.~3(i).\\
\indent\indent One would naively expect to see no slow phase dynamics beyond locking. Interestingly, however, Fig.~3(b) shows that the phase in both direct and parametric cases oscillates with a period of $\mathrm{\sim 0.1}$~ms. These oscillations are known as phase inertia \cite{barois2014frequency}. %, correspond to a situation where the phase of the oscillator behaves as an underdamped second order system, i.e. an underdamped resonator. % Similar oscillations were observed in SiC nanowire NEMS oscillators \cite{barois2014frequency}, but to our knowledge this the first such observation in a graphene oscillator.\\
 To extract the frequency of the phase oscillations, a Lorentzian  function is fitted to the PSD of the phase, as shown in the inset in Fig.~4(a). By fitting the PSD for the different drive powers at zero detuning, the dependence of the phase oscillation frequency on synchronization signal strength is obtained. Figure~4(a) shows these plotted on a logarithmic scale for both direct (blue) an parametric (red) entrainment. The frequency of the phase oscillation shows a power-law dependence on the strength of the reference signals. The exponents are $\mathrm{S_{d} = 0.56 \pm0.18}$ and $\mathrm{S_{p} = 0.61 \pm0.03}$, as obtained from the fits in Fig.~4(a).\\
\begin{figure}[h]
	\graphicspath{{Figures/}}
	\includegraphics[width=85mm]{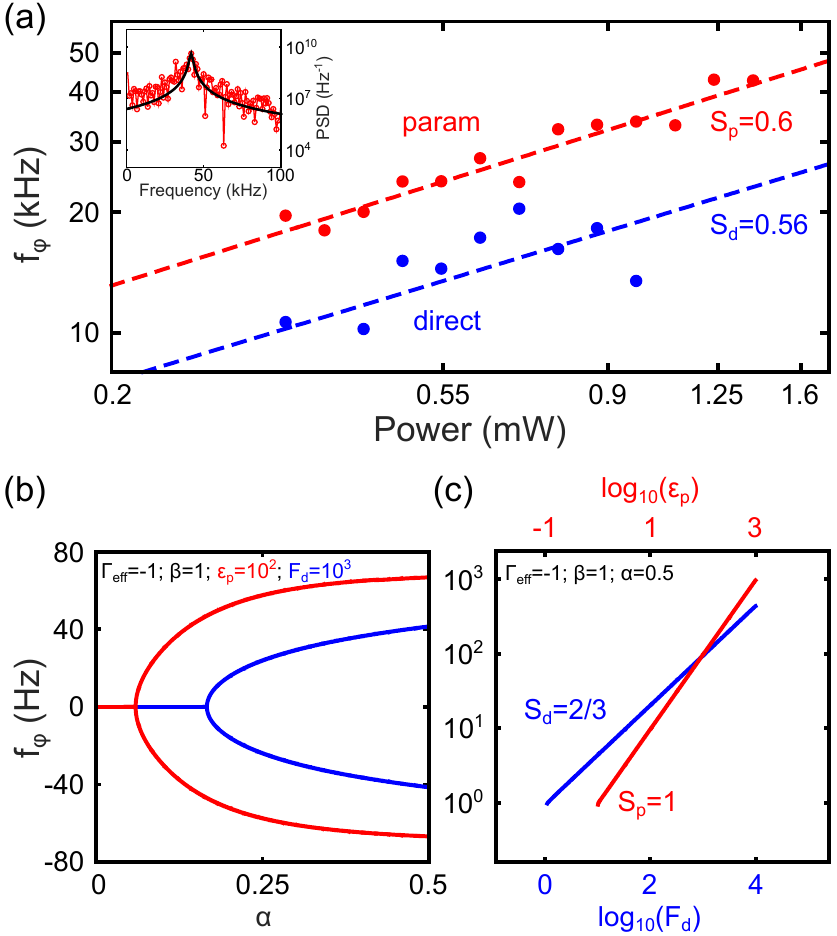}
	\caption{(a) Experimental power-law dependence of the phase resonance frequency on signal strength (rms signal power) for direct (blue) and parametric (red) locking. Inset shows a PSD of the phase and a Lorentzian fit. (b) Onset of phase oscillations as a function of $\mathrm{\pi}$ for direct (blue) and parametric (red) locking. (c) dependence of the phase oscillation frequency on direct (blue) and parametric (red) forcing. The parametric and direct cases present a linear and a sublinear dependence on forcing with $\mathrm{S_{d} = 2/3}$ and $\mathrm{S_{p} = 1}$}
\end{figure}
\indent\indent To capture the slow phase dynamics we model our system as a van der Pol oscillator with added terms to account for the Duffing nonlinearity, and the parametric and direct forcing \cite{nayfeh2008nonlinear}. The resulting forced van der Pol-Duffing-Mathieu equation expressed in non-dimensional form is given as : \\
\begin{equation}
\mathrm{\ddot x + ( \Gamma_{eff} + \beta x^2 ) \dot x + (1 -\varepsilon_p \cos ( \omega_p t )) x + \alpha x^3 = F_d \cos( \omega_d t)},
\end{equation}
\noindent
where the dot signifies taking the time-derivative, $\mathrm{x}$ is the normalized displacement, $\mathrm{\Gamma_{eff}}$ is the linear damping which in our case is negative due to photothermal feedback. $\mathrm{\beta}$ is a nonlinear damping term, $\mathrm{\varepsilon_p}$ is the strength of the parametric pumping term which is proportional to $\mathrm{P_{p}}$, $\mathrm{\omega_p}$ is the parametric pumping frequency, $\mathrm{\alpha}$ is the Duffing parameter, and $\mathrm{F_d}$ is the amplitude (proportional to $\mathrm{P_{d}}$) and $\mathrm{\omega_d}$ the frequency of the driving force. Note that for the cases studied in this work, the parametric forcing term $\mathrm{\varepsilon_p}$ and the direct forcing term $\mathrm{F_d}$ are never applied simultaneously. \\
\indent\indent The solution of Eq.~(2) is expressed in terms of a slowly changing phase $\mathrm{\phi(t)}$ and amplitude $\mathrm{A(t)}$, by taking $\mathrm{x(t) = A(t) \cos(\omega t + \phi(t))}$ \cite{balanov2008synchronization}. Such solutions have been reported for the forced van der Pol-Duffing-Mathieu equation in Refs.~\cite{pandey2008frequency,belhaq20082}. For zero detuning, Eq.~(2) can now be rewritten in terms of $\mathrm{\phi(t)}$ and $\mathrm{A(t)}$ as follows:\\
	\begin{eqnarray}
	\begin{rcases}
	\mathrm{\dot {\phi} = -\frac{\varepsilon_p}{2A} \sin(2\phi) + \frac{F_d}{2A} \cos(\phi) + \frac{3}{8}{\alpha}\mid{A}\mid^2}\\
	\mathrm{\dot {A} =  -\frac{\Gamma_{eff}}{2} - \frac{\beta}{8} \mid{A}\mid^2 A -\frac{\varepsilon_p}{2} \cos(2\phi) A - \frac{F_d}{2} \sin(\phi)}\\
	\end{rcases}
	\end{eqnarray}
\indent\indent Setting $\mathrm{\dot {\phi} = \dot {A} = 0}$ gives the stationary solution ($\mathrm{A_0}$) as follows: \\
	\begin{eqnarray}
	\mathrm{\left( \frac{9}{4} \bar{\alpha}^2 + \frac{\beta}{16} \right) A^6_0 + \frac{\Gamma_{eff}\beta}{2} A^4_0 + (\Gamma_{eff}^2 - \varepsilon_p^2)A^2_0 - {F_d^2} = 0.}
	\end{eqnarray}
\begin{comment}
\begin{subequations}
	\begin{eqnarray}
\begin{rcases}
\dot {\hat{\phi}} = \frac{3}{4} \bar{\alpha} A_0 \hat{A} - \left( \frac{\gamma}{2} + \frac{1}{8} A^2_0 \right) \hat{\phi}\\
\dot {\hat{A}} = - \left( \frac{\gamma}{2} + \frac{3}{8} A^2_0 \right) \hat{A} - \frac{3}{4} \bar{\alpha} A_0^3 \hat{\phi}\\
\end{rcases}
\\
\begin{rcases}
\dot {\hat{\phi}} = \frac{3}{4} \bar{\alpha} A_0 \hat{A} + \left( \frac{\gamma}{2} + \frac{1}{8} A^2_0 \right) \hat{\phi}\\
\dot {\hat{A}} = - \left( \frac{A^2_0}{4} \right) \hat{A} - \frac{3}{2} \bar{\alpha} A_0^3 \hat{\phi},
\end{rcases}
	\end{eqnarray}
\end{subequations}

\noindent
where the hat denotes a small deviation from stationary point, i.e.  $\hat{A}$ = $A$ - $A_0$ and $\hat{\phi}$ = $\phi$ - $\phi_0$, and $\bar{\alpha} = \alpha / 2\omega_0$, and the value of $A_0$ is obtained by solving the following polynomials for the direct and parametric cases respectively: \\
\begin{subequations}
	\begin{eqnarray}
	\left( \frac{9}{4} \bar{\alpha}^2 + \frac{1}{16} \right) A^6_0 + \frac{\gamma}{2} A^4_0 + \gamma^2 A^2_0 - \frac{F^2}{\omega^2_0} = 0 \\
	\left(\frac{9}{4} \bar{\alpha}^2 + \frac{1}{16} \right) A^4_0 + \frac{\gamma}{2} A^2_0 + \gamma^2 - \frac{\varepsilon^2}{\omega^2_0} = 0
	\end{eqnarray}
\end{subequations} %
\end{comment}
\indent\indent To study the slow phase dynamics we use a perturbative approach, where we set $\mathrm{\phi}$ = $\mathrm{\phi_0}$ + $\mathrm{\hat{\phi}}$, and $\mathrm{A}$ = $\mathrm{A_0}$ + $\mathrm{\hat{A}}$, with the hats denoting a small deviation from stationary solution. By inserting these into Eq.~(3), developing and keeping only first order terms, we obtain the following linear system of equations whose eigenvalues are the time constants of the phase oscillations:\\
	\begin{comment}
	\begin{widetext}
	\begin{eqnarray}
	\begin{rcases}
	\mathrm{\dot{\hat{\phi}} = -\left(\varepsilon_p \cos(2\phi_0) + \frac{F_d}{2A_0} \sin(\phi_0) \right)\hat{\phi} +\left(\frac{3}{4}{\alpha}A_0 + \frac{F_d}{2A_0^2} \cos(\phi_0) -  \frac{\varepsilon_p}{2A_0} \sin(2\phi_0) \right)\hat{A}} \\
	\mathrm{\dot{\hat{A}} =  -\left(\varepsilon_pA_0 \sin(2\phi_0) + \frac{F_d}{2} \cos(\phi_0)\right)\hat{\phi} + \left(\frac{\varepsilon_p}{2} \cos(2\phi_0)-\frac{\Gamma_{eff}}{2} - \frac{3}{8}\beta A_0^2\right)\hat{A}} \\
	\end{rcases}
	\end{eqnarray}
	\end{widetext}
	\end{comment}
		\begin{widetext}
			\begin{eqnarray}
			\begin{rcases}
			\mathrm{\dot{\hat{\phi}} = -\left(\varepsilon_p \cos(2\phi_0) + \frac{\Gamma_{eff}}{2} + \frac{1}{8}\beta A_0^2 \right)\hat{\phi} -\left(\frac{3}{4}{\alpha}A_0 +  \frac{\varepsilon_p}{2A_0} \sin(2\phi_0) \right)\hat{A}} \\
			\mathrm{\dot{\hat{A}} =  -\left(\varepsilon_pA_0 \sin(2\phi_0) + \frac{3}{4}\alpha A_0^3\right)\hat{\phi} + \left(\frac{\varepsilon_p}{2} \cos(2\phi_0)-\frac{\Gamma_{eff}}{2} - \frac{3}{8}\beta A_0^2\right)\hat{A}} \\
			\end{rcases}
			\end{eqnarray}
		\end{widetext}
\indent\indent The imaginary part of the eigenvalues of Eq.~(5) gives the phase resonance frequency $\mathrm{f_{\phi}}$. These are obtained and plotted in Fig.~4(b) as a function of the Duffing parameters (for $\mathrm{\Gamma_{eff} = -1}$, $\mathrm{\beta = 1}$, $\mathrm{\epsilon_p = 10^2}$, and $\mathrm{F_d = 10^3}$). For small $\mathrm{\alpha}$, the eigenvalues take only real values, indicating non-oscillatory, i.e. overdamped phase dynamics. As $\mathrm{\alpha}$ is increased, the eigenvalues become complex which indicates the transition to oscillatory phase behaviour. Figure~4(c) shows the dependence of the phase oscillation frequency on the synchronization signal strength for $\mathrm{\alpha} = 0.5$. In the case of direct forcing (blue trace) the time constant shows a sublinear dependence on signal strength (slope = 2/3) while parametric forcing (red trace) exhibits a linear dependence (slope = 1). Remarkably, increasing $\mathrm{\alpha}$ or $\mathrm{\Gamma_{eff}}$ has no influence on these slopes. Thus, once phase oscillation sets in, its power-law exponent is independent of both nonlinearity and oscillation amplitude.\\
\indent\indent For	$\mathrm{m = n = 1}$, the experimentally obtained power-law dependence with $\mathrm{S_d= 0.56}$ is in good agreement with the calculated $\mathrm{S_d= 2/3}$. This is less the case for parametric synchronization, $\mathrm{m = 2n = 2}$, where the experimentally obtained value is $\mathrm{S_p= 0.6}$ while in simulations $\mathrm{S_p= 1}$. The discrepancy could indicate the presence of additional nonlinearity, which may originate from device asymmetry that is introduced, for instance, by wrinkles or a non-uniformly distributed residual strain \cite{davidovikj2016visualizing}. The demonstrated phase oscillations are expected to occur naturally in entrained graphene oscillators, since they are easily driven into the nonlinear regime \cite{wang2014dynamic}, and their dependence on the drive strength and detuning with respect to the coupled reference oscillator may be used to further characterize the devices, or in applications that require the sensing of externally applied forces or masses \cite{papariello2016ultrasensitive}.\\
%\indent\indent Furthermore, since 2D NEMS are easily driven into the nonlinear regime thanks to their extreme geometrical aspect-ratio \cite{wang2014dynamic}, it should be expected that any entrained graphene nanomechanical oscillator will show phase oscillation behaviour. The frequency of the phase oscillations depends strongly on the drive and detuning. Thus, measurement of phase oscillations could be used for sensing small variations in force, amplitude and mass of the nanoscale graphene drum \cite{papariello2016ultrasensitive}.\\%
% \indent\indent In the photo-thermal actuation mechanism employed the applied force is directly proportional to the light intensity, which in our setup is proportional to the square of the applied signal power. Thus, by multiplying the experimentally obtained slopes by 2 in order to convert from signal power to actuation force, we find $S_{Direct} = 0.56$ and $S_{Param.} = 0.6$ to be in good agreement with simulation results. The remaining discrepancy between simulation and experiment is indicative of additional system nonlinearities whose exact source has yet to be identified. \\ %
\indent\indent In summary, the current work demonstrates that graphene self-oscillators can be synchronized to both a direct and parametric external signal at low temperatures. It is shown that achieving entrainment can significantly reduce the width of the oscillation peak, thus allowing reduction of oscillator frequency fluctuations to produce stable nanoscale oscillating motion. In addition to phase-locking and noise induced phase-slips, we also observed phase resonance and found that its frequency exhibits a power-law dependence on the drive signal strength for both direct and parametric synchronization. These oscillations were qualitatively reproduced using a forced van der Pol-Duffing-Mathieu equation, with the Duffing nonlinearity playing a crucial role in making such behaviour possible. This work enables synchronized motion of a large number of graphene oscillators, even if their resonance frequencies are slightly different. Potential applications of synchronized oscillators include optoelectronic modulators, sound generators and oscillating sensors.\\
 \indent\indent\ The authors acknowledge financial support from the European Union's Seventh Framework Programme (FP7) under Grant Agreement $\mathrm{n{\circ}~318287}$, project LANDAUER. The research leading to these results has received funding from the European Union's Horizon 2020 research and innovation program under Grant Agreement $\mathrm{n{\circ}~649953}$ (Graphene Flagship), and from the Dutch Technology Foundation STW Take-Off program, project $\mathrm{n{\circ}~14062}$.

\bibliographystyle{apsrev}
\bibliography{entrainment}

\begin{thebibliography}{25}
\expandafter\ifx\csname natexlab\endcsname\relax\def\natexlab#1{#1}\fi
\expandafter\ifx\csname bibnamefont\endcsname\relax
  \def\bibnamefont#1{#1}\fi
\expandafter\ifx\csname bibfnamefont\endcsname\relax
  \def\bibfnamefont#1{#1}\fi
\expandafter\ifx\csname citenamefont\endcsname\relax
  \def\citenamefont#1{#1}\fi
\expandafter\ifx\csname url\endcsname\relax
  \def\url#1{\texttt{#1}}\fi
\expandafter\ifx\csname urlprefix\endcsname\relax\def\urlprefix{URL }\fi
\providecommand{\bibinfo}[2]{#2}
\providecommand{\eprint}[2][]{\url{#2}}

\bibitem[{\citenamefont{Huygens}(1895)}]{huygens1895letters}
\bibinfo{author}{\bibfnamefont{C.}~\bibnamefont{Huygens}},
  \emph{\bibinfo{title}{Letters to de sluse,(letters; no. 1333 of 24 february
  1665, no. 1335 of 26 february 1665, no. 1345 of 6 march 1665)}}
  (\bibinfo{year}{1895}).

\bibitem[{\citenamefont{Oliveira and Melo}(2015)}]{oliveira2015huygens}
\bibinfo{author}{\bibfnamefont{H.~M.} \bibnamefont{Oliveira}} \bibnamefont{and}
  \bibinfo{author}{\bibfnamefont{L.~V.} \bibnamefont{Melo}},
  \bibinfo{journal}{Scientific reports} \textbf{\bibinfo{volume}{5}}
  (\bibinfo{year}{2015}).

\bibitem[{\citenamefont{Strogatz}(2003)}]{strogatz2003sync}
\bibinfo{author}{\bibfnamefont{S.~H.} \bibnamefont{Strogatz}},
  \emph{\bibinfo{title}{Sync: How order emerges from chaos in the universe,
  nature, and daily life}} (\bibinfo{publisher}{Hyperion},
  \bibinfo{year}{2003}).

\bibitem[{\citenamefont{Matheny et~al.}(2014)\citenamefont{Matheny, Grau,
  Villanueva, Karabalin, Cross, and Roukes}}]{matheny2014phase}
\bibinfo{author}{\bibfnamefont{M.~H.} \bibnamefont{Matheny}},
  \bibinfo{author}{\bibfnamefont{M.}~\bibnamefont{Grau}},
  \bibinfo{author}{\bibfnamefont{L.~G.} \bibnamefont{Villanueva}},
  \bibinfo{author}{\bibfnamefont{R.~B.} \bibnamefont{Karabalin}},
  \bibinfo{author}{\bibfnamefont{M.}~\bibnamefont{Cross}}, \bibnamefont{and}
  \bibinfo{author}{\bibfnamefont{M.~L.} \bibnamefont{Roukes}},
  \bibinfo{journal}{Physical review letters} \textbf{\bibinfo{volume}{112}},
  \bibinfo{pages}{014101} (\bibinfo{year}{2014}).

\bibitem[{\citenamefont{Bagheri et~al.}(2013)\citenamefont{Bagheri, Poot, Fan,
  Marquardt, and Tang}}]{bagheri2013photonic}
\bibinfo{author}{\bibfnamefont{M.}~\bibnamefont{Bagheri}},
  \bibinfo{author}{\bibfnamefont{M.}~\bibnamefont{Poot}},
  \bibinfo{author}{\bibfnamefont{L.}~\bibnamefont{Fan}},
  \bibinfo{author}{\bibfnamefont{F.}~\bibnamefont{Marquardt}},
  \bibnamefont{and} \bibinfo{author}{\bibfnamefont{H.~X.} \bibnamefont{Tang}},
  \bibinfo{journal}{Physical review letters} \textbf{\bibinfo{volume}{111}},
  \bibinfo{pages}{213902} (\bibinfo{year}{2013}).

\bibitem[{\citenamefont{Pikovsky et~al.}(2003)\citenamefont{Pikovsky,
  Rosenblum, and Kurths}}]{pikovsky2003synchronization}
\bibinfo{author}{\bibfnamefont{A.}~\bibnamefont{Pikovsky}},
  \bibinfo{author}{\bibfnamefont{M.}~\bibnamefont{Rosenblum}},
  \bibnamefont{and} \bibinfo{author}{\bibfnamefont{J.}~\bibnamefont{Kurths}},
  \emph{\bibinfo{title}{Synchronization: a universal concept in nonlinear
  sciences}}, vol.~\bibinfo{volume}{12} (\bibinfo{publisher}{Cambridge
  university press}, \bibinfo{year}{2003}).

\bibitem[{\citenamefont{Cross et~al.}(2004)\citenamefont{Cross, Zumdieck,
  Lifshitz, and Rogers}}]{cross2004synchronization}
\bibinfo{author}{\bibfnamefont{M.}~\bibnamefont{Cross}},
  \bibinfo{author}{\bibfnamefont{A.}~\bibnamefont{Zumdieck}},
  \bibinfo{author}{\bibfnamefont{R.}~\bibnamefont{Lifshitz}}, \bibnamefont{and}
  \bibinfo{author}{\bibfnamefont{J.}~\bibnamefont{Rogers}},
  \bibinfo{journal}{Physical review letters} \textbf{\bibinfo{volume}{93}},
  \bibinfo{pages}{224101} (\bibinfo{year}{2004}).

\bibitem[{\citenamefont{Shim et~al.}(2007)\citenamefont{Shim, Imboden, and
  Mohanty}}]{shim2007synchronized}
\bibinfo{author}{\bibfnamefont{S.-B.} \bibnamefont{Shim}},
  \bibinfo{author}{\bibfnamefont{M.}~\bibnamefont{Imboden}}, \bibnamefont{and}
  \bibinfo{author}{\bibfnamefont{P.}~\bibnamefont{Mohanty}},
  \bibinfo{journal}{Science} \textbf{\bibinfo{volume}{316}},
  \bibinfo{pages}{95} (\bibinfo{year}{2007}).

\bibitem[{\citenamefont{Zhang et~al.}(2012)\citenamefont{Zhang, Wiederhecker,
  Manipatruni, Barnard, McEuen, and Lipson}}]{zhang2012synchronization}
\bibinfo{author}{\bibfnamefont{M.}~\bibnamefont{Zhang}},
  \bibinfo{author}{\bibfnamefont{G.~S.} \bibnamefont{Wiederhecker}},
  \bibinfo{author}{\bibfnamefont{S.}~\bibnamefont{Manipatruni}},
  \bibinfo{author}{\bibfnamefont{A.}~\bibnamefont{Barnard}},
  \bibinfo{author}{\bibfnamefont{P.}~\bibnamefont{McEuen}}, \bibnamefont{and}
  \bibinfo{author}{\bibfnamefont{M.}~\bibnamefont{Lipson}},
  \bibinfo{journal}{Physical review letters} \textbf{\bibinfo{volume}{109}},
  \bibinfo{pages}{233906} (\bibinfo{year}{2012}).

\bibitem[{\citenamefont{Barois et~al.}(2014)\citenamefont{Barois, Perisanu,
  Vincent, Purcell, and Ayari}}]{barois2014frequency}
\bibinfo{author}{\bibfnamefont{T.}~\bibnamefont{Barois}},
  \bibinfo{author}{\bibfnamefont{S.}~\bibnamefont{Perisanu}},
  \bibinfo{author}{\bibfnamefont{P.}~\bibnamefont{Vincent}},
  \bibinfo{author}{\bibfnamefont{S.~T.} \bibnamefont{Purcell}},
  \bibnamefont{and} \bibinfo{author}{\bibfnamefont{A.}~\bibnamefont{Ayari}},
  \bibinfo{journal}{New Journal of Physics} \textbf{\bibinfo{volume}{16}},
  \bibinfo{pages}{083009} (\bibinfo{year}{2014}).

\bibitem[{\citenamefont{Barton et~al.}(2012)\citenamefont{Barton, Storch,
  Adiga, Sakakibara, Cipriany, Ilic, Wang, Ong, McEuen, Parpia
  et~al.}}]{barton2012photothermal}
\bibinfo{author}{\bibfnamefont{R.~A.} \bibnamefont{Barton}},
  \bibinfo{author}{\bibfnamefont{I.~R.} \bibnamefont{Storch}},
  \bibinfo{author}{\bibfnamefont{V.~P.} \bibnamefont{Adiga}},
  \bibinfo{author}{\bibfnamefont{R.}~\bibnamefont{Sakakibara}},
  \bibinfo{author}{\bibfnamefont{B.~R.} \bibnamefont{Cipriany}},
  \bibinfo{author}{\bibfnamefont{B.}~\bibnamefont{Ilic}},
  \bibinfo{author}{\bibfnamefont{S.~P.} \bibnamefont{Wang}},
  \bibinfo{author}{\bibfnamefont{P.}~\bibnamefont{Ong}},
  \bibinfo{author}{\bibfnamefont{P.~L.} \bibnamefont{McEuen}},
  \bibinfo{author}{\bibfnamefont{J.~M.} \bibnamefont{Parpia}},
  \bibnamefont{et~al.}, \bibinfo{journal}{Nano lett.}
  \textbf{\bibinfo{volume}{12}}, \bibinfo{pages}{4681} (\bibinfo{year}{2012}).

\bibitem[{\citenamefont{Metzger and Karrai}(2004)}]{metzger2004cavity}
\bibinfo{author}{\bibfnamefont{C.~H.} \bibnamefont{Metzger}} \bibnamefont{and}
  \bibinfo{author}{\bibfnamefont{K.}~\bibnamefont{Karrai}},
  \bibinfo{journal}{Nature} \textbf{\bibinfo{volume}{432}},
  \bibinfo{pages}{1002} (\bibinfo{year}{2004}).

\bibitem[{\citenamefont{Metzger et~al.}(2008)\citenamefont{Metzger, Favero,
  Ortlieb, and Karrai}}]{metzger2008optical}
\bibinfo{author}{\bibfnamefont{C.}~\bibnamefont{Metzger}},
  \bibinfo{author}{\bibfnamefont{I.}~\bibnamefont{Favero}},
  \bibinfo{author}{\bibfnamefont{A.}~\bibnamefont{Ortlieb}}, \bibnamefont{and}
  \bibinfo{author}{\bibfnamefont{K.}~\bibnamefont{Karrai}},
  \bibinfo{journal}{Phys. Rev. B} \textbf{\bibinfo{volume}{78}},
  \bibinfo{pages}{035309} (\bibinfo{year}{2008}).

\bibitem[{\citenamefont{Bunch et~al.}(2007)\citenamefont{Bunch, Van Der~Zande,
  Verbridge, Frank, Tanenbaum, Parpia, Craighead, and
  McEuen}}]{bunch2007electromechanical}
\bibinfo{author}{\bibfnamefont{J.~S.} \bibnamefont{Bunch}},
  \bibinfo{author}{\bibfnamefont{A.~M.} \bibnamefont{Van Der~Zande}},
  \bibinfo{author}{\bibfnamefont{S.~S.} \bibnamefont{Verbridge}},
  \bibinfo{author}{\bibfnamefont{I.~W.} \bibnamefont{Frank}},
  \bibinfo{author}{\bibfnamefont{D.~M.} \bibnamefont{Tanenbaum}},
  \bibinfo{author}{\bibfnamefont{J.~M.} \bibnamefont{Parpia}},
  \bibinfo{author}{\bibfnamefont{H.~G.} \bibnamefont{Craighead}},
  \bibnamefont{and} \bibinfo{author}{\bibfnamefont{P.~L.}
  \bibnamefont{McEuen}}, \bibinfo{journal}{Science}
  \textbf{\bibinfo{volume}{315}}, \bibinfo{pages}{490} (\bibinfo{year}{2007}).

\bibitem[{\citenamefont{Castellanos-Gomez
  et~al.}(2013)\citenamefont{Castellanos-Gomez, van Leeuwen, Buscema, van~der
  Zant, Steele, and Venstra}}]{castellanos2013single}
\bibinfo{author}{\bibfnamefont{A.}~\bibnamefont{Castellanos-Gomez}},
  \bibinfo{author}{\bibfnamefont{R.}~\bibnamefont{van Leeuwen}},
  \bibinfo{author}{\bibfnamefont{M.}~\bibnamefont{Buscema}},
  \bibinfo{author}{\bibfnamefont{H.~S.} \bibnamefont{van~der Zant}},
  \bibinfo{author}{\bibfnamefont{G.~A.} \bibnamefont{Steele}},
  \bibnamefont{and} \bibinfo{author}{\bibfnamefont{W.~J.}
  \bibnamefont{Venstra}}, \bibinfo{journal}{Advanced Materials}
  \textbf{\bibinfo{volume}{25}}, \bibinfo{pages}{6719} (\bibinfo{year}{2013}).

\bibitem[{\citenamefont{Chen et~al.}(2013)\citenamefont{Chen, Lee, Deshpande,
  Lee, Lekas, Shepard, and Hone}}]{chen2013graphene}
\bibinfo{author}{\bibfnamefont{C.}~\bibnamefont{Chen}},
  \bibinfo{author}{\bibfnamefont{S.}~\bibnamefont{Lee}},
  \bibinfo{author}{\bibfnamefont{V.~V.} \bibnamefont{Deshpande}},
  \bibinfo{author}{\bibfnamefont{G.-H.} \bibnamefont{Lee}},
  \bibinfo{author}{\bibfnamefont{M.}~\bibnamefont{Lekas}},
  \bibinfo{author}{\bibfnamefont{K.}~\bibnamefont{Shepard}}, \bibnamefont{and}
  \bibinfo{author}{\bibfnamefont{J.}~\bibnamefont{Hone}},
  \bibinfo{journal}{Nat. nanotechnol.} \textbf{\bibinfo{volume}{8}},
  \bibinfo{pages}{923} (\bibinfo{year}{2013}).

\bibitem[{\citenamefont{Adler}(1946)}]{adler1946study}
\bibinfo{author}{\bibfnamefont{R.}~\bibnamefont{Adler}},
  \bibinfo{journal}{Proceedings of the IRE} \textbf{\bibinfo{volume}{34}},
  \bibinfo{pages}{351} (\bibinfo{year}{1946}).

\bibitem[{\citenamefont{Paciorek}(1965)}]{paciorek1965injection}
\bibinfo{author}{\bibfnamefont{L.}~\bibnamefont{Paciorek}},
  \bibinfo{journal}{Proceedings of the IEEE} \textbf{\bibinfo{volume}{53}},
  \bibinfo{pages}{1723} (\bibinfo{year}{1965}).

\bibitem[{\citenamefont{Nayfeh and Mook}(2008)}]{nayfeh2008nonlinear}
\bibinfo{author}{\bibfnamefont{A.~H.} \bibnamefont{Nayfeh}} \bibnamefont{and}
  \bibinfo{author}{\bibfnamefont{D.~T.} \bibnamefont{Mook}},
  \emph{\bibinfo{title}{Nonlinear oscillations}} (\bibinfo{publisher}{John
  Wiley \& Sons}, \bibinfo{year}{2008}).

\bibitem[{\citenamefont{Balanov et~al.}(2008)\citenamefont{Balanov, Janson,
  Postnov, and Sosnovtseva}}]{balanov2008synchronization}
\bibinfo{author}{\bibfnamefont{A.}~\bibnamefont{Balanov}},
  \bibinfo{author}{\bibfnamefont{N.}~\bibnamefont{Janson}},
  \bibinfo{author}{\bibfnamefont{D.}~\bibnamefont{Postnov}}, \bibnamefont{and}
  \bibinfo{author}{\bibfnamefont{O.}~\bibnamefont{Sosnovtseva}},
  \emph{\bibinfo{title}{Synchronization: from simple to complex}}
  (\bibinfo{publisher}{Springer Science \& Business Media},
  \bibinfo{year}{2008}).

\bibitem[{\citenamefont{Pandey et~al.}(2008)\citenamefont{Pandey, Rand, and
  Zehnder}}]{pandey2008frequency}
\bibinfo{author}{\bibfnamefont{M.}~\bibnamefont{Pandey}},
  \bibinfo{author}{\bibfnamefont{R.~H.} \bibnamefont{Rand}}, \bibnamefont{and}
  \bibinfo{author}{\bibfnamefont{A.~T.} \bibnamefont{Zehnder}},
  \bibinfo{journal}{Nonlinear Dynamics} \textbf{\bibinfo{volume}{54}},
  \bibinfo{pages}{3} (\bibinfo{year}{2008}).

\bibitem[{\citenamefont{Belhaq and Fahsi}(2008)}]{belhaq20082}
\bibinfo{author}{\bibfnamefont{M.}~\bibnamefont{Belhaq}} \bibnamefont{and}
  \bibinfo{author}{\bibfnamefont{A.}~\bibnamefont{Fahsi}},
  \bibinfo{journal}{Nonlinear Dynamics} \textbf{\bibinfo{volume}{53}},
  \bibinfo{pages}{139} (\bibinfo{year}{2008}).

\bibitem[{\citenamefont{Davidovikj et~al.}(2016)\citenamefont{Davidovikj, Slim,
  Cartamil-Bueno, van~der Zant, Steeneken, and
  Venstra}}]{davidovikj2016visualizing}
\bibinfo{author}{\bibfnamefont{D.}~\bibnamefont{Davidovikj}},
  \bibinfo{author}{\bibfnamefont{J.~J.} \bibnamefont{Slim}},
  \bibinfo{author}{\bibfnamefont{S.~J.} \bibnamefont{Cartamil-Bueno}},
  \bibinfo{author}{\bibfnamefont{H.~S.} \bibnamefont{van~der Zant}},
  \bibinfo{author}{\bibfnamefont{P.~G.} \bibnamefont{Steeneken}},
  \bibnamefont{and} \bibinfo{author}{\bibfnamefont{W.~J.}
  \bibnamefont{Venstra}}, \bibinfo{journal}{Nano letters}
  \textbf{\bibinfo{volume}{16}}, \bibinfo{pages}{2768} (\bibinfo{year}{2016}).

\bibitem[{\citenamefont{Wang and Feng}(2014)}]{wang2014dynamic}
\bibinfo{author}{\bibfnamefont{Z.}~\bibnamefont{Wang}} \bibnamefont{and}
  \bibinfo{author}{\bibfnamefont{P.~X.-L.} \bibnamefont{Feng}},
  \bibinfo{journal}{Appl. Phys. Lett.} \textbf{\bibinfo{volume}{104}},
  \bibinfo{pages}{103109} (\bibinfo{year}{2014}).

\bibitem[{\citenamefont{Papariello et~al.}(2016)\citenamefont{Papariello,
  Zilberberg, Eichler, and Chitra}}]{papariello2016ultrasensitive}
\bibinfo{author}{\bibfnamefont{L.}~\bibnamefont{Papariello}},
  \bibinfo{author}{\bibfnamefont{O.}~\bibnamefont{Zilberberg}},
  \bibinfo{author}{\bibfnamefont{A.}~\bibnamefont{Eichler}}, \bibnamefont{and}
  \bibinfo{author}{\bibfnamefont{R.}~\bibnamefont{Chitra}},
  \bibinfo{journal}{arXiv preprint arXiv:1603.07774}  (\bibinfo{year}{2016}).

\end{thebibliography}

\end{document}